\documentclass[pre,twocolumn,amssymb,amsmath]{revtex4-1}
\usepackage{graphicx}
\usepackage{color}
\usepackage{epsfig}
\usepackage{latexsym}
\usepackage{bm}
\usepackage{ulem}

\usepackage{color}
\definecolor{purple}{rgb}{0.5,0,0.5}
\definecolor{dgreen}{rgb}{0,0.7,0}

\begin{document}

\renewcommand{\ni}{{\noindent}}
\newcommand{\dprime}{{\prime\prime}}
\newcommand{\be}{\begin{equation}}
\newcommand{\ee}{\end{equation}}
\newcommand{\bea}{\begin{eqnarray}}
\newcommand{\eea}{\end{eqnarray}}
\newcommand{\nn}{\nonumber}
\newcommand{\bk}{{\bf k}}
\newcommand{\bQ}{{\bf Q}}
\newcommand{\q}{{\bf q}}
\newcommand{\s}{{\bf s}}
\newcommand{\bN}{{\bf \nabla}}
\newcommand{\bA}{{\bf A}}
\newcommand{\bE}{{\bf E}}
\newcommand{\bj}{{\bf j}}
\newcommand{\bJ}{{\bf J}}
\newcommand{\bs}{{\bf v}_s}
\newcommand{\bn}{{\bf v}_n}
\newcommand{\bv}{{\bf v}}
\newcommand{\la}{\langle}
\newcommand{\ra}{\rangle}
\newcommand{\dg}{\dagger}
\newcommand{\br}{{\bf{r}}}
\newcommand{\brp}{{\bf{r}^\prime}}
\newcommand{\bq}{{\bf{q}}}
\newcommand{\hx}{\hat{\bf x}}
\newcommand{\hy}{\hat{\bf y}}
\newcommand{\bS}{{\bf S}}
\newcommand{\cU}{{\cal U}}
\newcommand{\cD}{{\cal D}}
\newcommand{\bR}{{\bf R}}
\newcommand{\pll}{\parallel}
\newcommand{\sumr}{\sum_{\vr}}
\newcommand{\cP}{{\cal P}}
\newcommand{\cQ}{{\cal Q}}
\newcommand{\cS}{{\cal S}}
\newcommand{\ua}{\uparrow}
\newcommand{\da}{\downarrow}
\newcommand{\red}{\textcolor {red}}
\newcommand{\blu}{\textcolor {blue}}
\newcommand{\1}{{\oldstylenums{1}}}
\newcommand{\2}{{\oldstylenums{2}}}
\newcommand{\mDelta}{\varepsilon}
\newcommand{\m}{\tilde m}
\def\lsim {\protect \raisebox{-0.75ex}[-1.5ex]{$\;\stackrel{<}{\sim}\;$}}
\def\gsim {\protect \raisebox{-0.75ex}[-1.5ex]{$\;\stackrel{>}{\sim}\;$}}
\def\lsimeq {\protect \raisebox{-0.75ex}[-1.5ex]{$\;\stackrel{<}{\simeq}\;$}}
\def\gsimeq {\protect \raisebox{-0.75ex}[-1.5ex]{$\;\stackrel{>}{\simeq}\;$}}

\title{Current reversal in interacting colloids under time-periodic drive }

\author{Shubhashis Rana, Sanchari Goswami, Sakuntala Chatterjee and Punyabrata Pradhan}

\affiliation{Department of Theoretical Sciences, S. N. Bose
National Centre for Basic Sciences, Block - JD, Sector - III, Salt
Lake, Kolkata 700106, India}

\newcommand{\nwc}{\newcommand}
\nwc{\vs}{\vspace}
\nwc{\hs}{\hspace}
\nwc{\lw}{\linewidth}
%
%
\nwc{\pd}[2]{\frac{\partial #1}{\partial #2}}
\nwc{\zprl}[3]{Phys. Rev. Lett. ~{\bf #1},~#2~(#3)}
\nwc{\zpre}[3]{Phys. Rev. E ~{\bf #1},~#2~(#3)}
\nwc{\zpra}[3]{Phys. Rev. A ~{\bf #1},~#2~(#3)}
\nwc{\zjsm}[3]{J. Stat. Mech. ~{\bf #1},~#2~(#3)}
\nwc{\zepjb}[3]{Eur. Phys. J. B ~{\bf #1},~#2~(#3)}
\nwc{\zrmp}[3]{Rev. Mod. Phys. ~{\bf #1},~#2~(#3)}
\nwc{\zepl}[3]{Europhys. Lett. ~{\bf #1},~#2~(#3)}
\nwc{\zjsp}[3]{J. Stat. Phys. ~{\bf #1},~#2~(#3)}
\nwc{\zptps}[3]{Prog. Theor. Phys. Suppl. ~{\bf #1},~#2~(#3)}
\nwc{\zpt}[3]{Physics Today ~{\bf #1},~#2~(#3)}
\nwc{\zap}[3]{Adv. Phys. ~{\bf #1},~#2~(#3)}
\nwc{\zjpcm}[3]{J. Phys. Condens. Matter ~{\bf #1},~#2~(#3)}
\nwc{\zjpa}[3]{J. Phys. A: Math theor  ~{\bf #1},~#2~(#3)}

\begin{abstract}
 
Using molecular dynamics simulations, we study particle-transport in a system of interacting colloidal particles on a  ring, where the system is driven by a time-dependent external potential, moving along the ring. We consider two  driving protocols: (i) the external potential barrier moves with a uniform velocity $v$ along the ring, and (ii) it moves in discrete jumps with jump-length $l$ and waiting time $\tau$ with an effective velocity $v=l/\tau$. The time-averaged (dc) particle current, which always remains positive in case (i), interestingly reverses its direction in case (ii) upon tuning the particle-number density $\rho_0$ and the effective barrier velocity $v$. We also find a scaling form for the current in terms of number density, barrier velocity, barrier height and temperature of the system. 

\end{abstract}

\maketitle

\section{Introduction} 

Characterizing particle transport in driven interacting-particle systems is an important problem in statistical physics \cite{rei02, Hanggi-review, jul97, chou11}. Over the past decades, there has been considerable progress made in maneuvering colloidal particles using laser field, leading to new avenues of research in exploring transport in such systems \cite{lib1, lib2, link08}. Several experiments have been performed using single colloidal particle, trapped in an external optical potential (optical tweezer), to investigate fundamental aspects of driven systems, such as fluctuation-response relations \cite{wang02, blic06, blic07, bech12, cili09, cili10}. Indeed, understanding transport in a system of interacting colloids driven by a time-periodic external force is important in the context of driven fluids in general and can have applications in developing stochastic pumps \cite{Ast, sini09}, thermal ratchets \cite{rei02, jul97, Hanggi-EPL, tara03, tara08, quake05}, and in controlling particle-transport through a confined geometry \cite{Cecchi, hangi02, Ai, Hanggi-PRL2007, Reimann, Mohanty,sandor17} like nanopores and micro-fluidic devices, etc.

In the past, there have been several studies to explore transport properties of simple model systems, such as simple exclusion processes (SEP) on a periodic lattice, where hardcore particles diffuse on discrete lattice sites and are driven by a time-dependent potential or force field \cite{jain07, mar08, pradhan14, pradhan16, dhar11}, with total particle-number in the systems being conserved. The main motivation behind these studies was to address an important question whether a time-periodic force can generate a nonzero time-averaged, or dc current even when the average (spatial or temporal) force acting on the system is zero. Interestingly, not only the answer has been in the affirmative, but, in certain cases \cite{jain07,pradhan14,pradhan16}, it has also been observed that the current, quite remarkably, can flow in both directions, depending on the values of certain parameters in the systems. Indeed, for two different time-periodic potentials used in Refs. \cite{jain07} and \cite{pradhan14, pradhan16}, the current reversal was obtained by tuning particle density, periodicity of the potential and other parameters of the system.

One particular advantage of the above mentioned simple lattice models is that the models are analytically tractable, albeit on a perturbative or a mean-field level. However, one may enquire whether the results derived from these models would apply to more realistic systems, where particles move in continuum, instead of discrete lattice. 
It should be noted that lattice spacing, which introduces a length scale into the problem, may be a relevant parameter in achieving current reversal. Therefore, it is not clear if, without employing an additional length scale, the results for a system on a discrete lattice would survive in the corresponding system in continuum, where the lattice spacing goes to zero. 
A recent study in this direction, however without introduction of any such length scale in the system, considers interacting Brownian particles diffusing in the presence of a travelling-wave-like external potential, which varies sinusoidally in space and time \cite{dhar15}. It was reported that this system also supports a nonzero current. But, {\it unlike} the lattice model, the direction of the current does not get reversed. For all choices of driving frequency and wave-length of the travelling-wave potential and for all particle densities, the current always flows in a particular direction, i.e., in the direction of the travelling wave \cite{dhar15}.

In this paper, we show that, depending on the driving protocol, it is possible to have a current reversal even for particles moving in continuum.  In our system of interacting Brownian particles, diffusing in continuum and on a ring geometry, we consider an external drive in the form of a time-dependent, and {\it spatially localized}, external potential barrier, which moves with velocity $v$ along the ring. We consider two driving protocols: Case (i)- the potential barrier moves continuously with a uniform velocity $v$ or case (ii)- it moves discretely, with its peak jumping from one position to the  other with a jump length $l$ and a waiting time $\tau$, thus having an effective velocity $v=l/\tau$ in this case. Note that, in driving protocol (ii), we have essentially introduced an additional length scale, the jump length of the moving barrier, which plays a crucial role here. Indeed, as we demonstrate, the time-averaged dc particle current remains always positive in the first case, but reverses its direction upon tuning the particle-number density $\rho_0$ and the effective velocity $v$, in the second case.

\begin{figure}[htbp]
\vspace{0.5cm}
\begin{center}
\includegraphics[width=8cm]{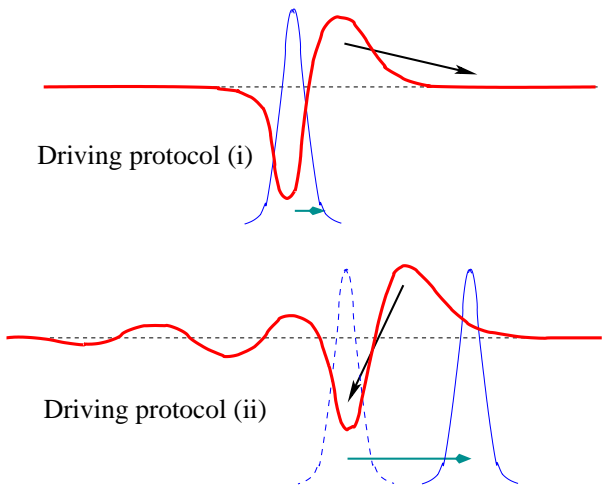}
\caption{ Schematic diagram of density profiles (red thick solid lines) for driving protocols (i) and (ii). In case (ii), the red line represents the density profile at the instant when the barrier has just moved to its new position (blue solid line) from its previous position (blue dashed line). The black arrows depict the gradient along which the dominant diffusive relaxation of the density profiles could occur. }
\label{schema1}
\end{center}
\end{figure}

The mechanism behind the current reversal, or rather the appearance of a negative current in the system, could be understood from diffusive relaxation of the density profile, which is created locally and is highly asymmetric around the barrier position; see the schematic diagrams in Fig. \ref{schema1}. For driving protocol (i), the barrier always stays behind the density peak, which is formed just in front of it. Therefore, the particle current is almost entirely generated by the diffusive relaxation of the density peak to the bulk on the right side of the barrier. In this case, the relaxation of the density peak to the density trough on the left is prohibited due to the presence of the barrier there. On the other hand, for driving protocol (ii), as the barrier keeps on jumping to a new position after a waiting time $\tau$, the density peak can now relax into the density trough to the left, giving rise to a significant negative current contribution to the total particle current. Moreover, the presence of the already shifted barrier in front of the density peak hinders the relaxation of density peak to the right, contributing to the reduction in the positive current. Therefore, the current reversal essentially arises from the competition between the above two (positive and negative) current contributions; for a quantitative description of the mechanism, see Fig. \ref{cur-comp}. Depending on the height(s) of the density peak(s), which again depends on the parameters $v$ and $\rho_0$, the net current can be either positive or negative.
The results obtained here nicely complement our earlier studies of a simple lattice model, where a current reversal was observed and explained using a similar mechanism described above \cite{pradhan14, pradhan16}.

The organization of the paper is as follows. In Sec. II, we define the model. We present the results for continuous barrier movement in section III and for discrete barrier movement in Sec. IV. We summarize and conclude in Sec. V.  

\begin{figure}[htbp]
\vspace{0.5cm}
\begin{center}
\includegraphics[width=8cm]{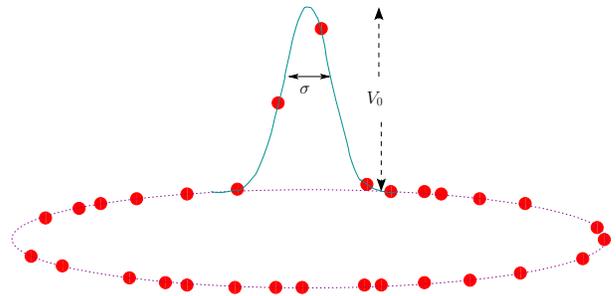}
\caption{Schematic diagrams of the model. A Gaussian potential barrier (shown in cyan color) moves along a one dimensional ring (represented as a pink dashed circle). The colloidal particles are represented as red balls, which cannot cross each other due to the divergence of the WCA potential $V_{WCA}(r)$ as $r$ approaches zero; however, slight overlap between two particles, though unlikely at low temperature, is dynamically allowed (soft core). }
\label{schema}
\end{center}
\end{figure}

\section{The Model}

We consider $N$ identical Brownian particles, diffusing along a one dimensional ring of length $L$. In the overdamped limit, we can neglect the inertial term; this assumption is quite reasonable for micron-sized colloids in a viscous medium as, in that case, the velocities of individual colloids would relax very fast due to large viscous drag and therefore one could ignore the acceleration, or the inertial term. Consequently, the equation of motion of the $i$th particle in the system can be written as,
\begin{equation}
 \dot x_i = - \beta D \left[ \sum_{j \ne i} \frac{\partial V_{WCA}(r_{ij})}{\partial x_i}  + \frac{\partial V(x_i,t)}{\partial x_i} \right] + \eta_i,
\label{L1}
\end{equation}
where $x_i$ is position of the $i$th particle, $r_{ij}=|x_j-x_i|$ is relative distance between particles $i$ and $j$. The interaction potential $V_{WCA} (r_{ij} )$  between $i$th and $j$th particle is chosen to be the Weeks-Chandler-Anderson (WCA) potential, which has the following functional form,
\begin{equation}
 V_{WCA}(r)= 4\epsilon\left[\left(\frac{a}{r} \right)^{12}-\left(\frac{a}{r} \right)^{6}\right] + \epsilon,
\end{equation}
for $r < 2^{1/6}a$ with $\epsilon=1$ and $a$ being the particle diameter, while $V_{WCA}(r)=0$ otherwise. The external potential $V(x,t)$ acting at position $x_i$ and time $t$, has a Gaussian form with a moving peak:
\begin{equation}
 V(x,t)= V_0 e^{- [x-x_0(t)]^2/2 \sigma^2},
\end{equation}
where $x_0(t)$ is the position of the center or the peak of the potential barrier at time $t$, $\sigma$ is the width of the barrier and $V_0$ is the strength (see Fig. \ref{schema}). In Eq. \ref{L1},  the fluctuating force term $\eta_i$ is considered to be a Gaussian white noise with vanishing mean force $\langle \eta_i \rangle=0$ and delta correlations $\langle \eta_i(t)\eta_j(t')\rangle =2 D\delta_{i,j}\delta(t-t')$, where diffusion constant $D$ and inverse temperature $\beta = 1/k_BT$ are related through fluctuation-dissipation relation $D= \mu k_B T$ with Boltzmann constant $k_B$ and mobility $\mu$ both being taken to be unity (thus $D=T$) throughout the paper.

We consider two driving protocols in the paper. (i) In the first case, the potential barrier moves {\it continuously} along the ring with a uniform speed $v$,
$$
x_0(t) = vt.
$$
(ii) In the second case, the potential barrier moves along the ring, by {\it discrete jumps}, from one position to another, by a jump length $l$ and then it waits at the new position for a residence time $\tau$, and so on, such that
$$
x_0(t) = \delta(x - n v \tau)
$$ 
with $n = 0, 1, 2, \dots, \infty$ and the effective velocity of the barrier is $v=l/\tau$. Clearly, in the limit of small jump length and waiting time $l, \tau \rightarrow 0$ with velocity $v=l/\tau$ fixed, we recover case (i) of continuous movement of the barrier.

Due to the periodic boundary condition, the system, in both the cases, eventually settles into a time-periodic steady state with time period $\tau_c = L/v$, time over which the barrier completes one cycle by circling around the ring once. Note that there is another time scale, a microscopic diffusive time $\tau_0=a^2/D$ over which a particle diffuses across a length scale of the particle diameter $a$. The global number density is defined as $\rho_0=N/L$. We measure length in the unit of $a$ and, therefore, put $a=1$ throughout. We also take the width of the potential barrier $\sigma=1$ and use system size $L=200$ in our study. 

To integrate Eq. \ref{L1} numerically, we use Heun's method \cite{heun}, where we discretize time in steps of $\delta t=10^{-4} \tau_0$ and calculate the position $x_i(t+ \delta t)$ of the $i$-th particle at time $t+ \delta t$ from its velocities $\dot x_i(t)$ and $\dot x_i^e(t+ \delta t)$, where $x_i^e(t+\delta t)$ is estimated using Euler's method. This integration technique yields second order accuracy ${\mathcal{O}}[(\delta t)^2]$.

\section{Continuous movement of the barrier}

\begin{figure}[!ht]
\vspace{0.5cm}
\begin{center}
 \includegraphics[width=8cm]{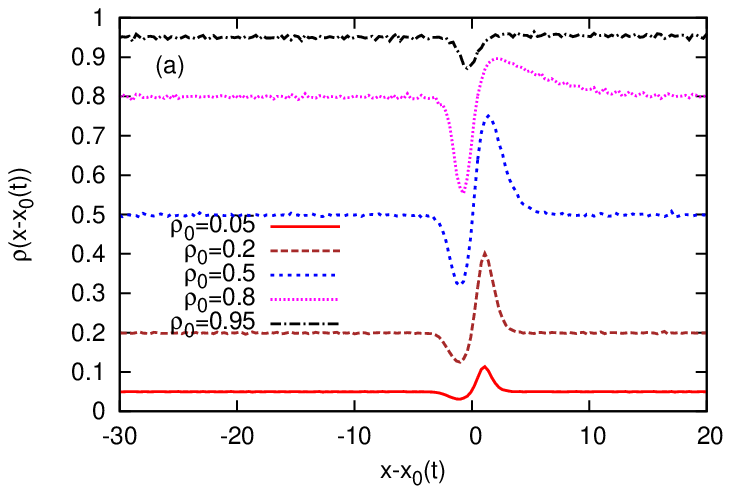}
\includegraphics[width=8cm]{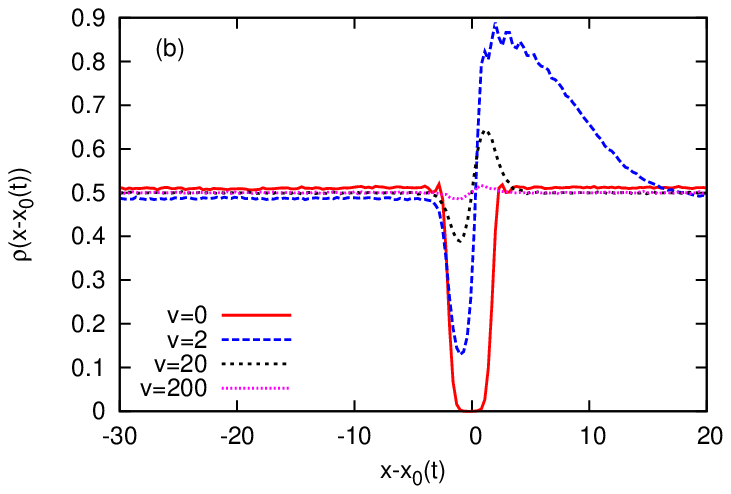}
\caption{ Particle density profiles for continuous movement of the barrier [driving protocol (i)]. (a) Local density profile $\rho(x-x_0(t))$ as a function of the distance from the barrier position $x_0(t)$, for different values of global density $\rho_0=0.05$(red continuous line), $0.2$(brown dashed line), $0.5$(blue dotted line), $0.8$(magenta fine dotted line) and $0.95$(black dash-dotted line). The topmost curve corresponds to highest $\rho_0$ and the lower curves correspond to decreasing $\rho_0$ values in succession. We have used $v=10$, $T=1$ and $V_0=10$. (b) Local density profile $\rho(x-x_0(t))$ for different barrier speeds $v=0$(red continuous line), $2$(blue dashed line), $20$(black dotted line) and $200$(magenta fine dotted line).  At $v=0$ (equilibrium), the density profile has only one localized trough, and is otherwise homogeneous. At very high velocity, the density profile tends to a homogeneous one as any point in the system does not have sufficient time to feel the barrier. The peak and the trough in the density profile become more pronounced in an intermediate range of $v$. Here we have used $\rho_0=0.5$, $T=1$ and $V_0=10$.}
\label{Number-density}
\end{center}
\end{figure}

\begin{figure}[!ht]
\vspace{0.5cm}
\begin{center}
 \includegraphics[width=8cm]{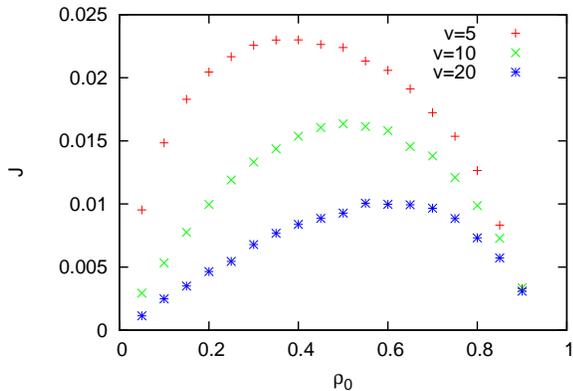}
\caption{Driving protocol (i). Particle current $J$ is plotted {\it vs.} global density $\rho_0$ for three different barrier velocities $v=5$(red pluses), $10$(green crosses), $20$(blue asterisks) while $V_0=10$ and $T=1.0$ are kept fixed.  }
\label{Number}
\end{center}
\end{figure}

In this section, we study case (i) where the external potential barrier moves along the ring continuously. Expectedly, in equilibrium when $v=0$, there is a trough (rarefied region with lower particle density compared to the bulk) in the density profile, with density value being minimum  exactly at the position of the barrier peak ( See Fig. \ref{Number-density}(b)). When the barrier moves with a nonzero velocity $v$, the system eventually reaches a time-periodic state with a density travelling wave moving along a ring with the same velocity $v$ and having a hump (compressed region with higher particle density) followed by a trough.  In this case of continuous movement of the barrier with velocity $v$, the density trough slightly lags behind the barrier peak.  In Fig. \ref{Number-density}(a), we plot local density as a function of the distance from the barrier position $x_0(t)$, for a fixed $v$ and various values of global density $\rho_0$.  
Due to the barrier movement, particles get accumulated in front of the barrier and get depleted at the back. This structure is very similar to that observed in the lattice version of the model previously studied in Ref. \cite{pradhan14}. In Fig. \ref{Number-density}(b) we plot the density profile  for different values of the barrier speed $v$. Our data show that, for smaller values of $v$, the trough and peaks in the density profile are more prominent and, as $v$ increases, the height (depth) of the peak (trough) becomes smaller.

We also measure the particle current, which is defined as the average number of particles flowing across a particular point on the ring per unit time. In Fig. \ref{Number}, we plot current $J$ as a function of global density $\rho_0$ for various values of the barrier velocity $v$. Here we find that, unlike in the lattice model of Ref. \cite{pradhan14}, there is no current reversal, i.e., current $J$ remains always positive. However, current $J$ is still a nonmonotonic function of density $\rho_0$. Initially, as $\rho_0$ increases, $J$ first increases, reaches its peak and then decreases as $\rho_0$ increases further. Notably, there is no particle-hole symmetry and therefore the particle current is not symmetric around $\rho_0=1/2$.

\begin{figure}[!ht]
\begin{center}
\includegraphics[width=8cm]{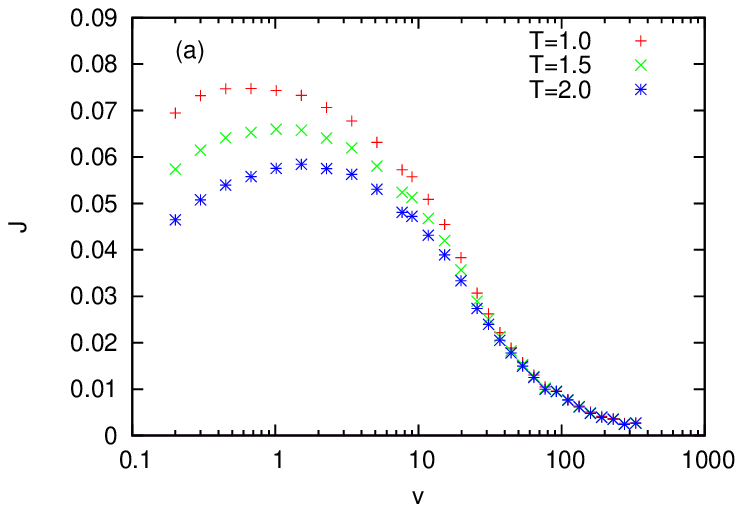}
\includegraphics[width=8cm]{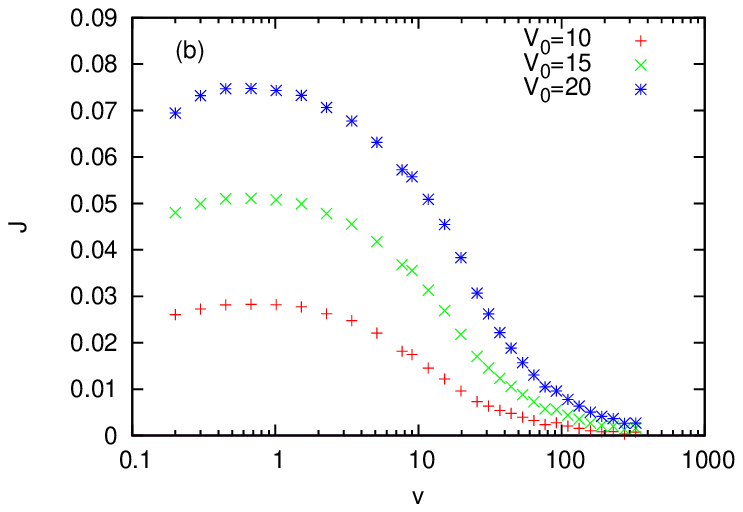}
\caption{ Driving protocol (i). (a) Current $J$ is plotted as a function of the barrier velocity $v$ for different temperatures $T=1$(red pluses), $1.5$(green crosses), $2.0$(blue asterisks) with strength of the barrier $V_0 = 20$ kept fixed.  (b) Current $J$ is plotted as a function of $v$ for $V_0=10$(red pluses), $15$(green crosses), $20$(blue asterisks) with temperature $T=1$ kept fixed. Here we have taken global density $\rho_0=0.5$.}
\label{con-cur-vel}
\end{center}
\end{figure}

Next we plot particle current $J$ as a function of barrier velocity $v$ in Fig. \ref{con-cur-vel}, for different values of $T$ and $V_0$. The current shows a peak at an intermediate velocity regime. From Fig. \ref{con-cur-vel}(a), we find that, as temperature increases with the barrier height $V_0$ kept fixed, the current decreases. On the other hand, if the barrier height $V_0$ decreases with the temperature kept fixed, the particle current starts decreasing, as shown in Fig. \ref{con-cur-vel}(b).

\begin{figure}[!ht]
\begin{center}
\includegraphics[width=8cm]{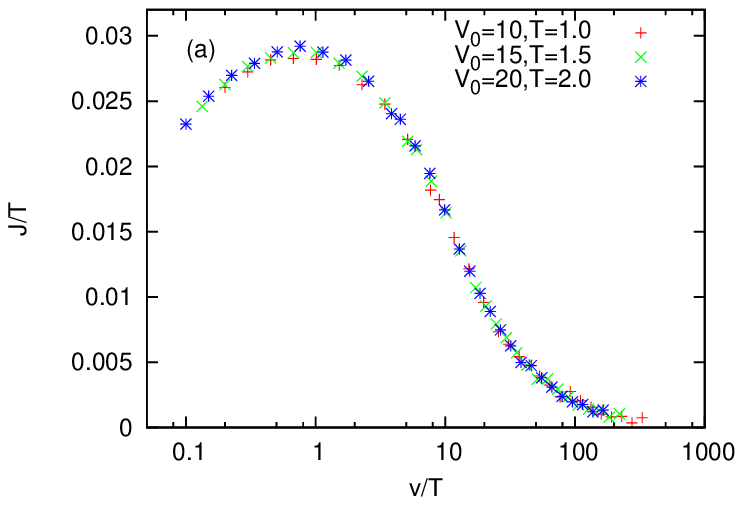}
\includegraphics[width=8cm]{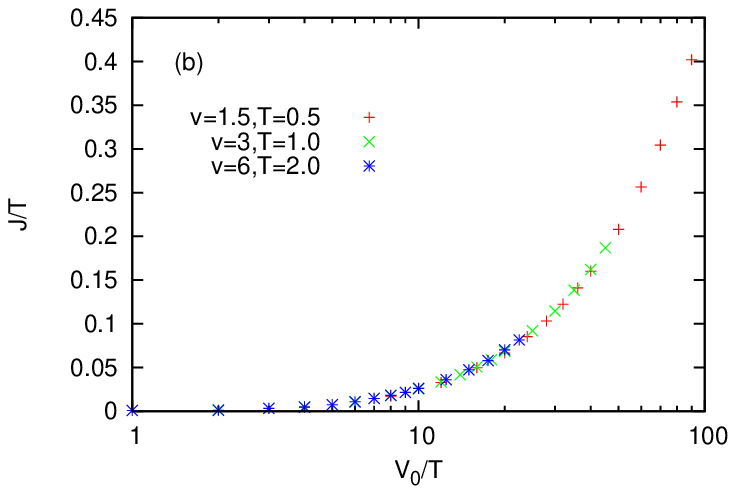}
\caption{Driving protocol (i). (a) Scaled particle current $J/T$ is plotted as a function of scaled barrier velocity, or Peclet number, ${\rm Pe} = v/T$ where $V_0/T=10$ is kept fixed. (b) $J/T$ is plotted as a function of $V_0/T$ for  $v/T =3$. }
\label{con-cur-vel2}
\end{center}
\end{figure}  

The particle current can be written as a function of the following variables - global density $\rho_0$, barrier velocity $v$, barrier strength $V_0$ and temperature $T$,
\be 
J = J(\rho_0, v, V_0, T).
\ee
Note that the arguments inside the above function must be dimensionless. Now, considering instantaneous local current $j(x,t) \simeq D \partial \rho(x)/\partial x$ being essentially diffusive, with $\partial \rho(x)/\partial x$ being local density gradient, the total time-averaged current $J \simeq \int_0^{\tau_c} dt \int_0^L dx j(x,t)/\tau_c$ would be proportional to diffusivity $D=T$ and we have the functional form $J/T = f(\rho_0, v \tau_0/a, V_0/T)$ for current, where the dimensionless variable ${\rm Pe}=v \tau_0/a$  is known as Peclet number. In other words, we have a scaling form,
\begin{equation}
\frac{J}{T} = f_1\left( \rho_0, \frac{v}{T},\frac{V_0}{T} \right),
\label{scaling1}
\end{equation}
which we verify from our data in Fig. \ref{con-cur-vel2}(a), where we plot 
scaled current $J/T$ as a function of scaled velocity, or Peclet number, ${\rm Pe}=v/T$ (using $a=1$) with the scaled barrier height $V_0/T$ kept fixed and observe a good scaling collapse. Similarly, in Fig. \ref{con-cur-vel2}(b) we plot the scaled current $J/T$ as a function of scaled potential $V_0 / T$ for fixed ${\rm Pe}$ and find a good collapse. In the above scaling calculation, we have neglected possible dependence of the diffusivity $D$ on local density. However, even when it depends on the local density, the diffusivity is expected to be still proportional to the temperature. The good scaling collapse shown by our data is consistent with this assumption.

Note that the particle current always remains positive for all ranges of the scaling variables. This is in line with the conclusion drawn for sinusoidally varying space-time dependent external potential studied in Ref. \cite{dhar15}. This observation brings us to the main question of this paper: Is it possible to find a driving protocol for a system in continuum such that there is current reversal upon tuning certain parameters of the system? In the next section, we answer this question in the affirmative.

\section{Discrete jump of the barrier}

In this section, we consider an external potential barrier, which is moving around the system in discrete steps with jump length $l$ and waiting time $\tau$. The waiting time $\tau$ is the residence time during which the barrier stays put at a particular position and then jumps instantaneously to another position, $l$ distance away, along a particular direction (say, counter-clockwise). Therefore, the barrier moves with an effective velocity $v=l/\tau$ along the ring with a time-period $\tau_c=L/v$. One could vary both $l$ and $\tau$, but in most of the cases we study here, we fix the jump length at $l=4$ (where, as discussed later, negative current contribution is significant) and vary waiting time $\tau$ to obtain data for different barrier velocities $v$.

\begin{figure}[!ht]
\vspace{0.5cm}
\begin{center}
 \includegraphics[width=8cm]{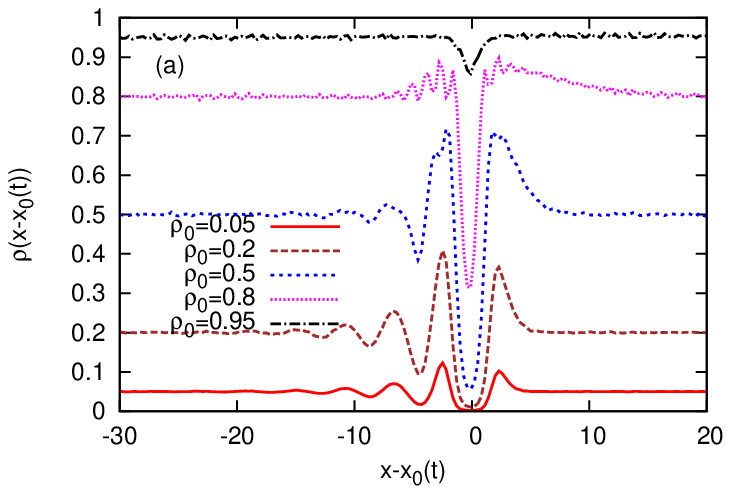}
\includegraphics[width=8cm]{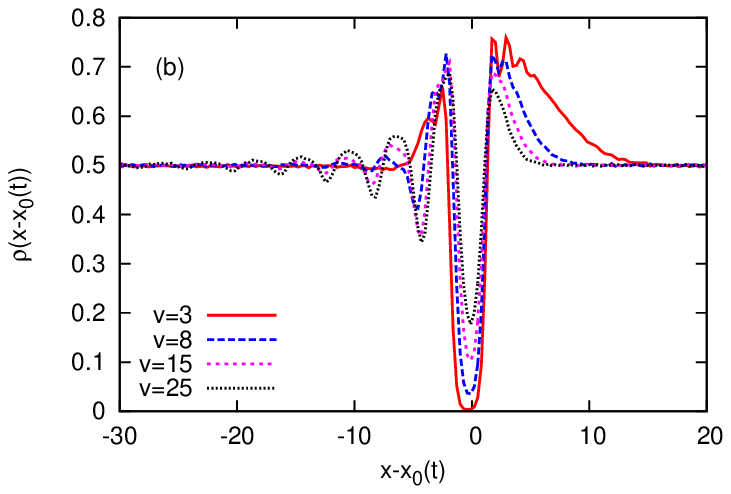}
\caption{Driving protocol (ii). (a) Local particle density $\rho(x-x_0(t))$ as function of distance from the barrier position $x_0(t)$, at a particular time when the center of the barrier is about to jump to the new location. The different global density  $\rho_0$ values are shown in the legends and we have used here  $v=10$. (b) Density profile $\rho(x-x_0(t))$  for different barrier velocity $v =3$(red continuous line), $8$(blue dashed line), $15$(magenta dotted line) and $25$(black fine dotted line). In all cases, temperature $T=1$ and barrier strength $V_0=10$.}
\label{Number-density2}
\end{center}
\end{figure}

The system reaches a time-periodic steady state with period $\tau_c$ and there appears a travelling density wave moving along the ring with a velocity $v$. In Fig. \ref{Number-density2}(a), we have plotted, for different values of global density $\rho_0$, the local particle density of the system as a function of distance from the barrier position, measured at a particular time when the center of the barrier is about to jump to the next location. The spatial structure of travelling wave is quite similar to that in Fig. \ref{Number-density} and in the previously studied lattice version of the model \cite{pradhan14}, except that now there are oscillations behind the center of the barrier over large length scales. The wave length of these oscillations is determined by the jump-length $l$. The magnitude of the oscillations is small when $\rho_0$ is too low or too high. For intermediate $\rho_0$ values, the oscillations are most pronounced. Unless stated otherwise, we work with $\rho_0 = 0.5$ from now on. To the right of the barrier position, there is a density peak, which is followed by a density trough at the center of the barrier. In Fig. \ref{Number-density2}(b) we plot the density profile for different values of barrier speed $v$ and find that, as $v$ increases, the peak and trough become less prominent since the barrier spends less time at a particular position for large $v$.  Even after the barrier has jumped to a new position, the peak and trough created around its old positions, persist for some time before diffusion homogenizes them. This gives rise to a trail of peak and trough pairs of varying magnitudes along the path of the barrier. This explains why the above mentioned oscillations in the density profile are observed only behind the barrier and not in front of it.

\begin{figure}[!ht]
\vspace{0.5cm}
\begin{center}
 \includegraphics[width=8cm]{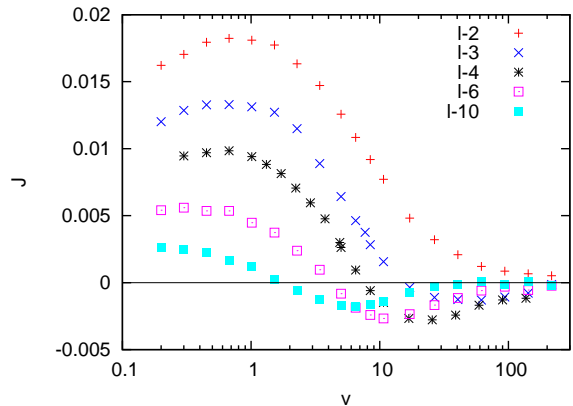}
\caption{Driving protocol (ii). Particle current $J$ is plotted as a function of effective barrier velocity $v=l/\tau$ by varying the waiting time $\tau$ for fixed values of jump-lengths $l$. The largest magnitude of negative current is observed for jump length $l=4$. Here, we have used barrier height $V_0=10$ and temperature $T=1$.}
\label{jump-length-ls} 
\end{center}
\end{figure}

To study how the particle current $J$ is affected by discrete jump of the potential barrier, we plot in Fig. \ref{jump-length-ls} the variation of $J$ with barrier speed $v$ for different values of the jump length $l$. As explained in the beginning of this section, we vary $v$ by holding $l$ fixed and changing $\tau$. Our data in Fig. \ref{jump-length-ls} show that, for small values of $l$, current remains positive for all $v$. This is consistent with our finding in the previous section, since one expects to retrieve the results for continuous barrier movement in the limit of small $l$. However, as $l$ increases, we find $J$ becomes negative for a certain $v$ range, before vanishing for large $v$. In other words, for particles moving in continuum, acted upon by an external potential moving in discrete jumps, the particle current $J$ shows a positive peak for smaller $v$ values and a negative peak for larger $v$ values, just as found in \cite{pradhan14,pradhan16} for a lattice model. Our data also shows that the negative peak of $J$ is most pronounced for $l=4$. For jump lengths much larger than this, the overall magnitude of $J$ goes significantly down and the positive as well as negative peaks become rather small. Since we are mainly interested in the negative part of $J$, in the remaining part of this section, we keep $l$ fixed at the value $4$, and investigate how other parameters in the system affect the negative $J$.

\begin{figure}[!ht]
\vspace{0.5cm}
\begin{center}
 \includegraphics[width=8cm]{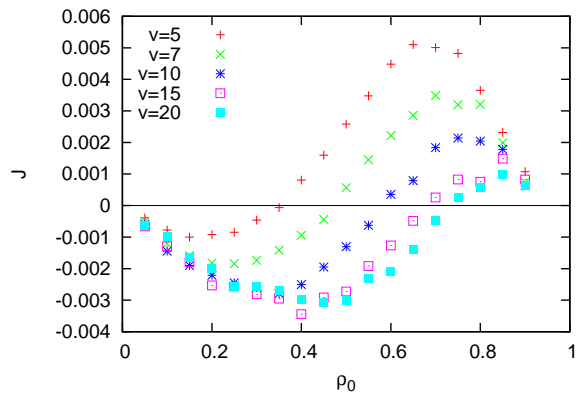}
\caption{Driving protocol (ii). Particle current $J$ is plotted as a function of global density $\rho_0$ for different barrier velocities $v=5$(red pluses), $7$(green crosses), $10$(blue asterisks), $15$(magenta open boxes) and $20$(cyan filled boxes), with temperature $T=1$ and barrier strength $V_0=10$ fixed.}
\label{J-rho2}
\end{center}
\end{figure}

\begin{figure}[!ht]
\vspace{0.5cm}
\begin{center}
\includegraphics[width=8cm]{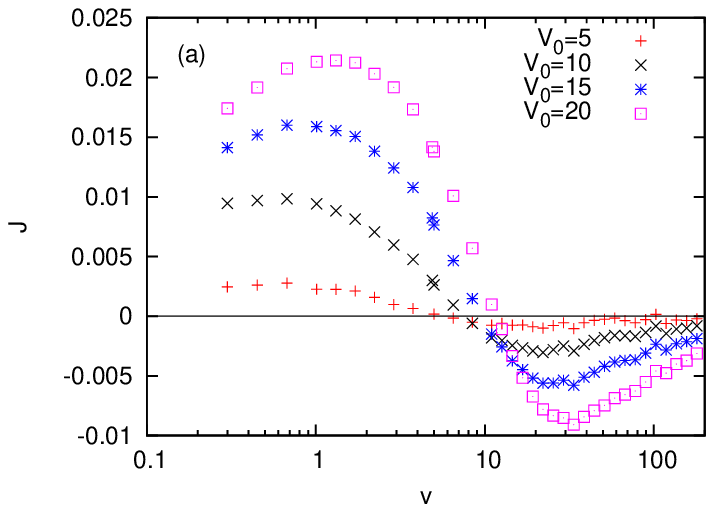}
\includegraphics[width=8cm]{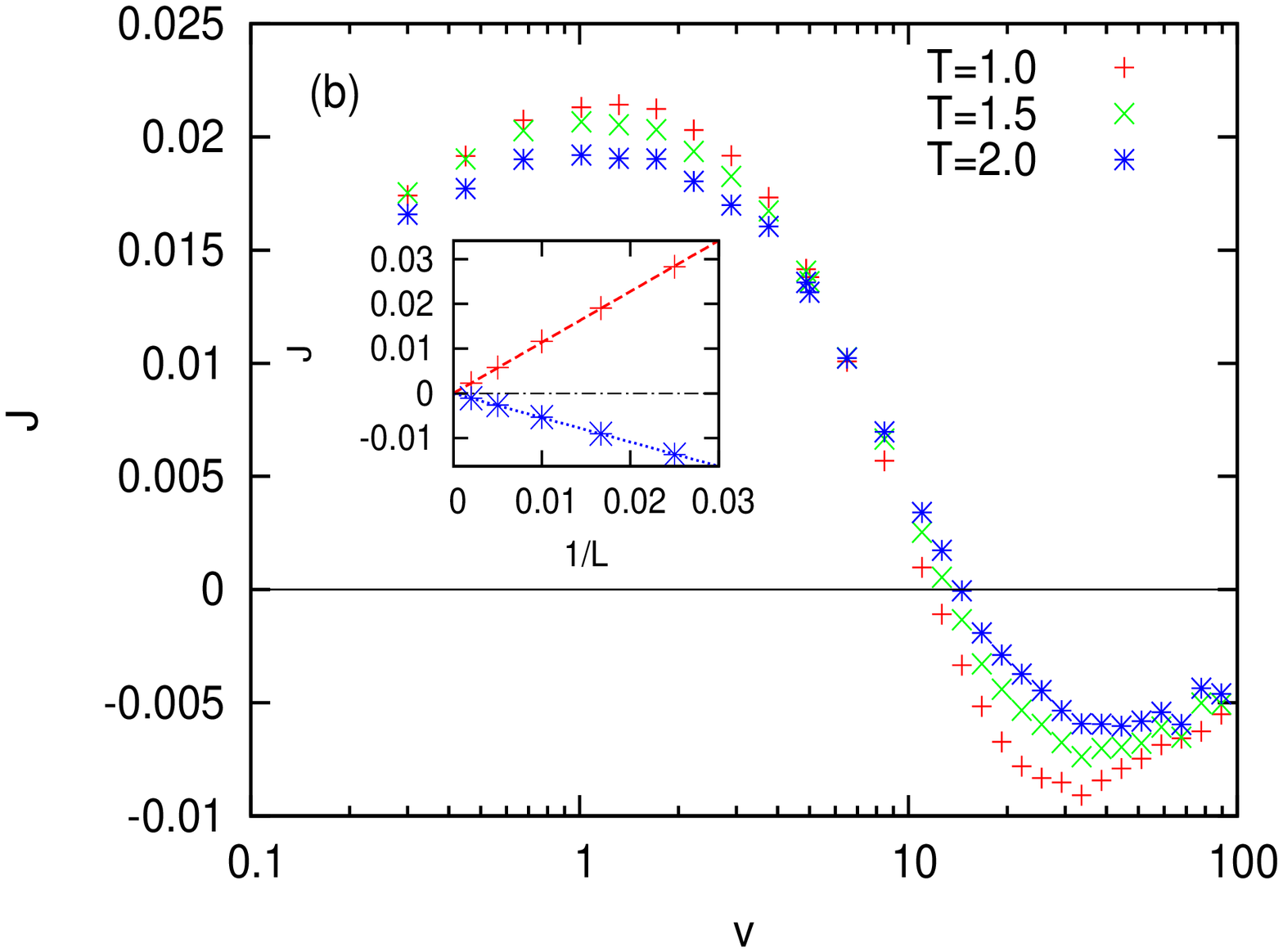}
\caption{Driving protocol (ii). Particle current $J$ is plotted as a function of barrier velocity. In panel (a), we have used different values of barrier strengths $V_0=5$(red pluses), $10$(black crosses), $15$(blue asterisks) and $20$(magenta open boxes) and $T=1$. In panel (b), the curves correspond to different temperatures $T=1$(red pluses), $1.5$(green crosses) and $2$(blue asterisks) while $V_0=20.0$. In all cases, we take global density $\rho_0=0.5$ and jump-length $l=4$. In the inset of panel (b), we show $J \sim 1/L$ scaling for $v=3$(red plus points) and $v=25$ (blue asterisk points) taking $T=1.0$ fixed.} 
\label{jump-length} 
\end{center}
\end{figure}

In Fig. \ref{J-rho2} we show the variation of particle current $J$ as a function of particle density $\rho_0$ for different values of $v$. Negative $J$ is observed for smaller values of $\rho_0$. The current shows a negative peak, followed by a positive peak, whose positions and height depend strongly on $v$. For a fixed jump length $l=4$, as $v$ is increased by decreasing the residence time $\tau$, we find that the negative peak shifts to higher values of $\rho_0$ and its height also increases, whereas the positive peak becomes less and less pronounced for large $v$.  In Figs. \ref{jump-length}, we plot $J$ as a function of $v$ for different potential strengths $V_0$ [panel (a)] and different temperatures $T$ [panel  (b)]. Both the panels show that the particle current has a positive peak, followed by a negative peak. Not surprisingly, the height of these peaks and indeed the overall magnitude of the current decreases (increases) as the temperature (potential strength) increases. In the inset of Fig. \ref{jump-length}(b), we plot the current as a function of the inverse of the system size $L$. Clearly, the current decays as $J \sim 1/L$, which is expected as the current is generated locally around the barrier, through density relaxation, and the barrier comes back at a particular point after a time period $L/v$. Most importantly, our data in Figs. \ref{J-rho2} and \ref{jump-length} clearly show that, when the global number density $\rho_0$ is small and/or the barrier velocity $v$ is large, the system supports a negative current. 

\begin{figure}[!ht]
\begin{center}
\includegraphics[width=8cm]{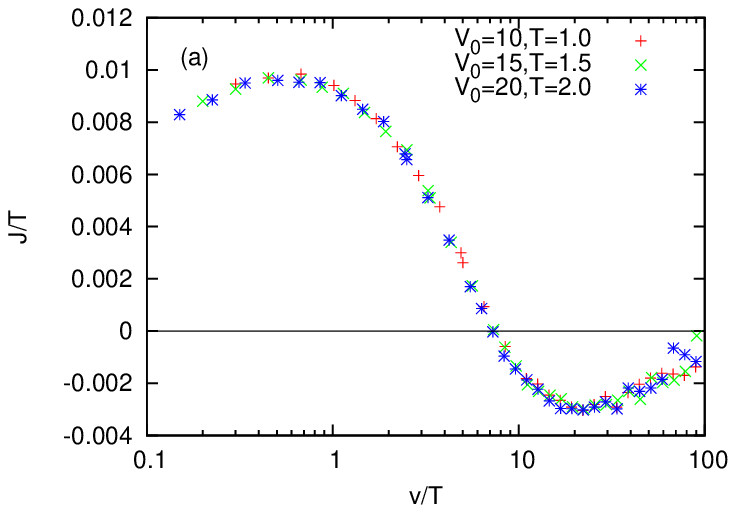}
\includegraphics[width=8cm]{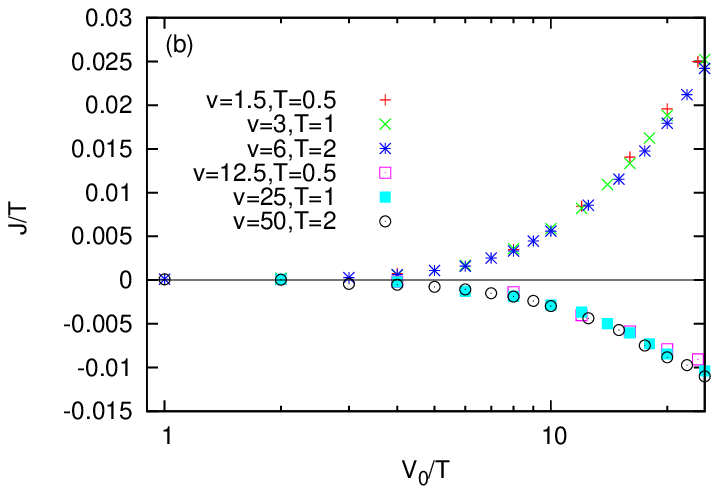}
\caption{Driving protocol (ii). (a) Scaled current $J/T$ is plotted as a function of scaled barrier velocity $v/T$ for different values of barrier strength $V_0$ and temperature $T$ with keeping $V_0/T=10$ fixed. (b) Scaled current $J/T$  is plotted as a function of scaled barrier strength $V_0/T$ for fixed $v/T$. Upper curve is for $v/T =3$ and the lower curve is for $v/T=25$. }
\label{cur-vel}
\end{center}
\end{figure}

The dimensional analysis presented in the previous section regarding the scaling properties of $J$, remains valid in the case of discrete barrier movement as well. Therefore, we expect the following scaling form for $J$
\be 
\frac{J}{T} = f_2 \left( \rho_0, \frac{v}{T},\frac{V_0}{T} \right),
\label{scaling2}
\ee
where the scaling function $f_2$ is naturally expected to be different from $f_1$ appearing in Eq.\ref{scaling1}. In the top panel of Fig. \ref{cur-vel}(a) 
we plot scaled current $J/T$ as a function of scaled barrier velocity $v/T$, keeping scaled barrier strength $V_0/T$ fixed. We find a quite good scaling collapse. In  Fig. \ref{cur-vel}(b), we plot $J/T$ as a function of $V_0/T$ by keeping $v/T$ fixed at two different values. As expected, the sign of current is negative when $v/T$ takes a large value. 

As mentioned before, the particle current is essentially diffusive in nature and can be understood from the structure of the density profile as in our previously studied lattice model in Ref. \cite{pradhan14, pradhan16}. To validate the mechanism behind the current reversal, as described in the introduction, we study density relaxation by tagging particles behind and in front of the barrier. During the waiting time $\tau$, the barrier remains static and the contributions to positive (anti-clockwise) and negative (clockwise) currents arise mainly due to the movement of particles to the right and the left of the peaks along the downhill slopes of the density profile, i.e., from the peaks to the troughs. Below we explicitly study the space-time trajectories of several particles around the barrier and explicitly measure the current contributions to the right and the left of the barrier.

\begin{figure}[!ht]
\vspace{0.5cm}
\begin{center}
 \includegraphics[width=8.5cm]{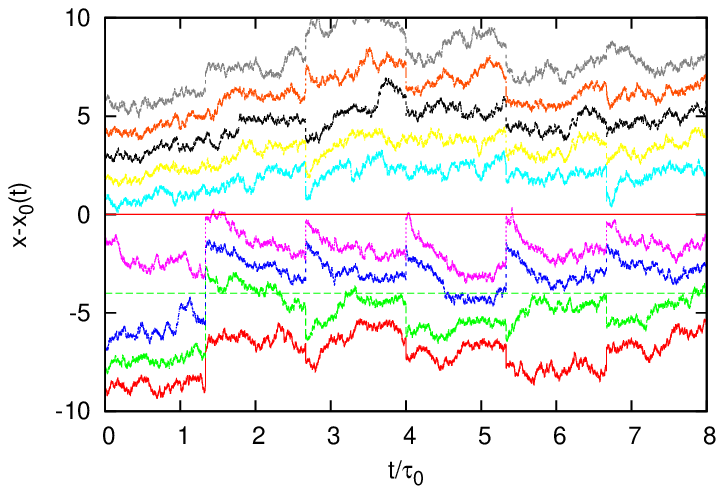}
 \includegraphics[width=8.5cm]{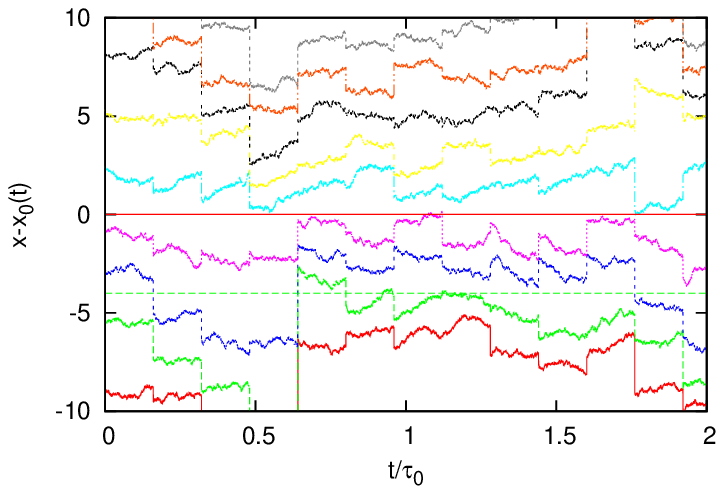}
\caption{Driving protocol (ii). The space-time trajectories of particles around the barrier $x_0(t)$. Different colors have been used to show the trajectories of different particles. The tags of the particles are changed, each time the barrier jumps. Here we have used  $v=3$ (top panel) and  $v=25$ (bottom panel). The other parameters are set at $l=4$, $V_0=10$, $T=1.0$ for both plots. $\tau_0=a^2/D$ is diffusive time unit.}
\label{traj}
\end{center}
\end{figure}

\begin{figure}[!ht]
\vspace{0.5cm}
\begin{center}
\includegraphics[width=8cm]{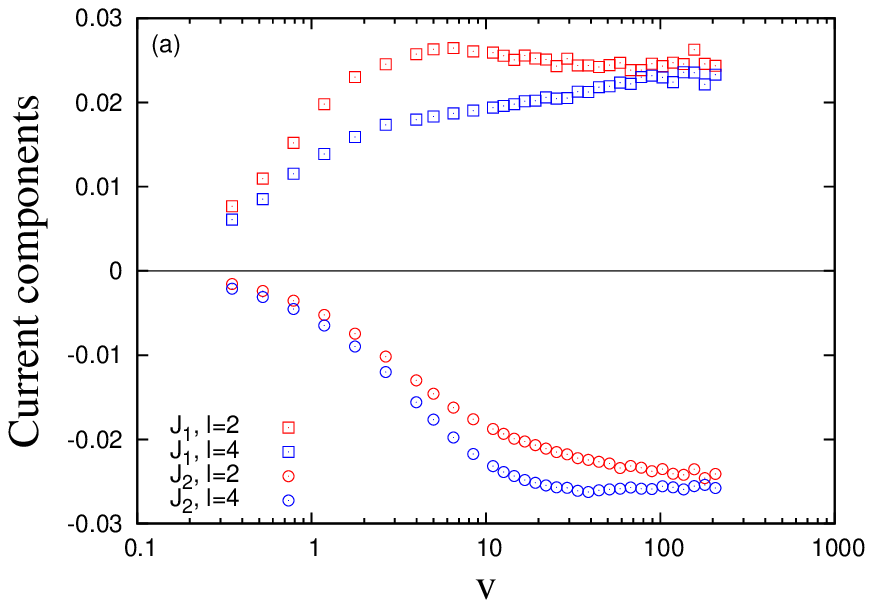}
\includegraphics[width=8cm]{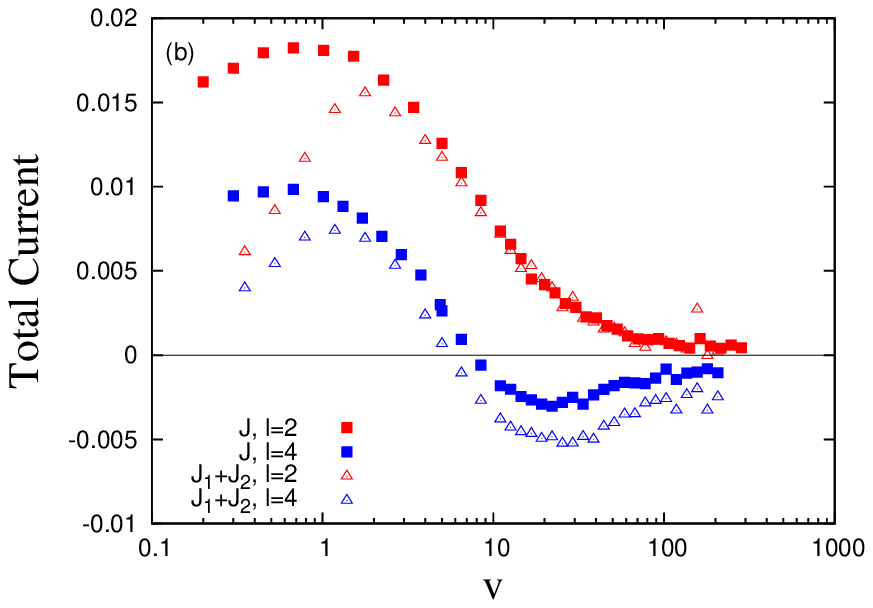}
\caption{Driving protocol (ii). (a) The positive and negative components of the current as a function of barrier velocity. The red points correspond to jump length $l=2$ and blue points correspond to $l=4$. The positive current $J_1$ is shown by box-symbols and the negative current $J_2$ by circles. (b) Comparison between total current $J$ and sum of $J_1 + J_2$ as a function of $v$. The red points are for $l=2$ and blue points are for $l=4$. The solid squares represent total $J$ and hollow triangles represent $J_1 + J_2$.}
\label{cur-comp} 
\end{center}
\end{figure}

The top panel in Fig. \ref{traj} corresponds to a slow barrier movement ($v=3$) and the bottom panel corresponds to a fast barrier movement ($v=25$). Every-time the barrier moves, the tag of each particle is changed such that for each position of the barrier, the trajectories of all particles within a certain distance from the barrier are shown. We have used different colors to depict the trajectories of different tagged 
particles. As seen from these trajectories, the particles which are to the right of the barrier, show a net rightward displacement and those on the left side of the barrier show a net leftward displacement. This is expected since the particles always tend to move away from the potential barrier. In other words, contribution to the particle current is positive (negative) for those particles on the right (left) of the barrier. To verify this, we explicitly measure current $J_1$ ($J_2$) averaged over a region on the right-side (left-side) of the barrier. In Fig.  \ref{cur-comp}(a) we plot $J_1$ and $J_2$ as a function of $v$ and, as expected, $J_1$ is always positive and $J_2$ remains always negative. The total current $J$ is approximately the sum of these two contributions. In the bottom panel of Fig. \ref{cur-comp}, we compare $J_1 + J_2$ with the total current $J$ and find qualitatively quite similar behavior between the two quantities. The observed difference between the two quantities is presumably due to the averaging over particles in a limited region around the barrier. From the above analysis, it follows therefore that the negative current arises when magnitude of $J_2$ exceeds that of $J_1$. Comparing with the density profile as in Fig. \ref{Number-density2}(b), we also observe that the negative contribution $J_2$ takes a large magnitude if the jump-length $l$ is such that, after each jump, the barrier lands just at the right of the density peak that was created around its old position. This is expected since there are  more particles contributing to $J_2$ in this situation.  The excess density, which is now at immediate left of the new barrier position, will relax by filling up the density trough created at the old barrier position and, due to large density gradient present in this region, the corresponding diffusive current (which is negative in this case) will have a large magnitude, leading to a net negative current in the system.

This mechanism is quite similar to the one in the lattice model studied in Ref. \cite{pradhan14}, for which current was calculated analytically within a mean-field theory. However, an analytical  calculation
in a continuum system, especially when the barrier jumps in discrete time steps, as considered in this paper, 
turns out to be more challenging.

\section{Summary and concluding remarks} 

In this paper, we have studied a set of driven colloidal particles interacting with each other via short ranged potential and experiencing an externally applied potential barrier that moves around the system with a fixed (average) velocity. We find that the presence of an external time-periodic drive gives rise to a travelling wave density profile in the system that moves with the same velocity as that of the potential barrier. However, the particle current that flows through the system depends strongly on the specific protocol of barrier movement. When the barrier moves continuously with velocity $v$, the current always flows in the direction of the barrier movement. But when the barrier moves through the system in discrete jumps, then it is possible to have current flowing in the opposite direction. In both cases we find a scaling form for the current as a function of barrier height and barrier speed, scaled by system temperature.

In a recent study \cite{dhar15}, a similar system was considered where colloidal particles under a sinusoidally 
varying travelling wave potential were shown to support a current that only flows in the direction of the travelling
wave. Although a lattice model version of the system was studied in Ref. \cite{jain07} where negative
current was  found, it was concluded in Ref. \cite{dhar15}  that, for particles moving in continuum, 
it may not be possible to have a negative current. But our present work shows that, even in continuum, 
a negative current can be obtained, if the potential moves in discrete jumps in the system. 
Our study demonstrates that a naive extension of a lattice model to continuum may not always reproduce the qualitative features of particle transport observed in the original lattice model. This is because the lattice spacing implicitly introduces a length scale, which could be important in the problem as the phenomenon of current reversal happens due to the local diffusive relaxation of density. In our work, by using the driving protocol (ii), we incorporate an additional length scale in the system in the form of a jump length of the moving barrier. We have demonstrated that the jump length indeed plays a crucial role in the phenomenon of current reversal.
The conclusions of this paper, which supports the mechanism of current reversal described above, are also consistent with our earlier work on a lattice model studied in Ref. \cite{pradhan14,pradhan16}.

Finally, it would be quite exciting to  verify some of our conclusions in experiments. Recently, it has been possible to experimentally realize a mono-layer of interacting colloidal particles in a quasi-periodic potential \cite{bech12, bechin17} whose amplitude was periodically modulated in time. The motion of the colloidal particles was studied in the presence of a driving force acting laterally on the mono-layer and interesting effects like dynamical ordering \cite{bech12}, shapiro-step like dependence of particle velocity on applied force \cite{bechin17} has been observed. Our conclusions can be tested in a similar experimental set up where colloidal particles are trapped in a one dimensional channel and then, by using the driving protocol discussed in this paper, it would be quite interesting to see if a current reversal can be observed in a real system. 

%
%
%

\section{Acknowledgement}

The computational facility used in this work was provided through the Thematic Unit of Excellence (TUE) on Computational Materials Science, funded by Nanomission, Department of Science and Technology (DST), India. S.R. thanks the Department of Science and Technology (DST), India for the financial support through National Postdoctoral Fellowship (NPDF), Science and Engineering Research Board (SERB), India.

\end{document}